\documentclass[12pt]{iopart}

\usepackage{iopams}

\usepackage[nohl,end]{optional}

\usepackage[T1]{fontenc}
\usepackage[latin1]{inputenc}
\usepackage{color,graphicx}
\usepackage{ifpdf}
\usepackage{amsmath,amsfonts}
\usepackage{hyperref}

\hypersetup{
pdfauthor = {Noé Curtz},
pdftitle = {Patterning of ultrathin YBCO nanowires using a new focused-ion-beam process},
pdfsubject = {},
pdfkeywords = {},
pdfcreator = {LaTeX / hyperref},
pdfproducer = {latex}
}

\ifpdf
\DeclareGraphicsExtensions{.pdf,.png}
\else
\DeclareGraphicsExtensions{.ps,.eps}
\fi

\opt{hlvtwo}{ %
\newcommand{\vtwo}[1]{{\color{blue}{#1}}}
}

\opt{nohl}{ %
\newcommand{\vtwo}[1]{#1}
}

\opt{end}{
\def\fgwidth{\textwidth}
\def\iswidth{5cm} %
\def\mbl{44pt} %
\def\rbl{144pt} %
}
\opt{inline}{
\def\fgwidth{.5\textwidth}
\def\iswidth{2.5cm}
\def\mbl{22pt}
\def\rbl{72pt}
}

\begin{document}

\title{Patterning of ultrathin YBCO nanowires using a new focused-ion-beam process}

\author{N~Curtz$^{1,2}$, E~Koller$^2$, H~Zbinden$^1$, M~Decroux$^2$, L~Antognazza$^2$, {\O}~Fischer$^2$ and N~Gisin$^1$}
\address{$^1$ Group of Applied Physics, University of Geneva, 1211, Geneva 4, Switzerland}
\address{$^2$ Department of Condensed Matter Physics, University of Geneva, 1211, Geneva 4, Switzerland}
\ead{Noe.Curtz@unige.ch}

\begin{abstract}
Manufacturing superconducting circuits out of ultrathin films is a challenging task when it comes to pattern complex compounds, which are likely to be deteriorated by the patterning process. With the purpose of developing high-T$\text{c}$ superconducting photon detectors, we designed a novel route to pattern ultrathin YBCO films down to the nanometric scale. \vtwo{We believe that our method, based on a specific use of a focused ion beam, consists in locally implanting Ga$^\text{3+}$ ions and/or defects instead of etching the film. This protocol} could be of interest to engineer high-Tc superconducting devices (SQUIDS, SIS/SIN junctions, Josephson junctions), as well as to treat other sensitive compounds.
\end{abstract}


\maketitle
\date{\today}

\section{Introduction}

Superconducting Single-Photon Detectors (SSPDs) are superconducting devices designed with the purpose of detecting light down to single-photon level. They present a good quantum efficiency (QE > 10\%), low dark-count rate (DK < 10 Hz), and high operating frequency (> GHz), outperforming in a number of cases InGaAs Avalanche PhotoDiodes (APDs) \cite{Thew-NIMA-2009}. These characteristics make them a premium candidate for single-photon telecommunication and applications like Quantum Key Distribution. Their underlying mechanism is based upon the formation of a hotspot in a current-biased superconducting stripe \cite{Goltsman-APL-2001}. The creation of the hotspot is triggered by the incoming photon whose energy locally thermalizes the stripe, confining the bias current hence raising its density, up to the point of overcoming its critical value, resulting in a local transition of the stripe.

To achieve that, the circuit geometry has to fulfill drastic geometrical constraints. First, the stripe has to be narrow enough, otherwise the variation of the current density isn't sufficiently significant, preventing the transition to take place and the voltage pulse to be detectable. The section of the nanowire should also be extremely homogeneous, since any constriction locally lowers the critical current, hence affects the whole device with for aftermath a drop of the QE; the closer to 1 is the ratio $I_\text{bias}/I_\text{c}$, the closer to the dissipative state is the device, and therefore the more important is the section homogeneity. Finally, the detector's recovery time is governed by the device thickness, necessitating devices less than 15 nm thick. Whereas such devices have been successfully produced with low-Tc superconductors such as NbN operated at 2.6 K \cite{Marsili-OE-2008}, their realization with high-Tc compounds remains a challenge. High-T$\text{c}$ SSPDs would nevertheless allow a higher working temperature, hence a significant reduction of the associated cryogenic costs.

Among high-T$_\text{c}$ materials, cuprates present the advantage of a low kinetic inductance, leading to fast response times. From a purely structural point of view, Nd$_{\text{1+}x}$Ba$_{\text{2-}x}$Cu$_\text{3}$O$_{\text{7+}\delta}$ presents excellent crystallographic and planeity properties \cite{Badaye-SST-1997}, which are interesting features given the aforementioned geometrical constraints of SSPDs. It turns out however that the intrinsic loose stoichiometry of Nd atoms, who interdiffuse with Ba ones, leads to a high unstability of the oxygen content and therefore to an important loss of T$_\text{c}$ during the patterning process. YBa$_\text{2}$Cu$_\text{3}$O$_{\text{7-}\delta}$ (YBCO) is much more stable and appears to be a good candidate for high-T$_\text{c}$ detectors. Several routes have been carried out to create junctions or patterned structures out of \vtwo{high-T$_\text{c}$} thin films, such as Selective Epitaxial Growth \cite{Damen-SST-1998}, Electron Beam Lithography / Ion Beam Etching (EBL/IBE) \cite{Schneidewind-PHYSC-1995}, Ion Irradiation \cite{Bergeal-APL-2006}, using an Atomic Force Microscope \cite{Delacour-APL-2007} or a Focused Electron Beam Irradiation \cite{Booij-PRB-1997}. Nanobridges have been fabricated with a Focused Ion Beam \cite{Lee-PHYSC-2007}. However those experiments were performed on films with thickness $d$ > 20 nm. Here we report a new method using such an apparatus to write an arbitrary pattern upon an ultrathin (< 20 nm) film, allowing to manufacture YBCO superconducting circuits. The key point of this method is that to produce the structure the superconducting phase is locally altered rather than etched.

\section{Experimental}

\subsection{Overview of the FIB-based protocol}

\begin{figure}[p]
\includegraphics[width=\fgwidth]{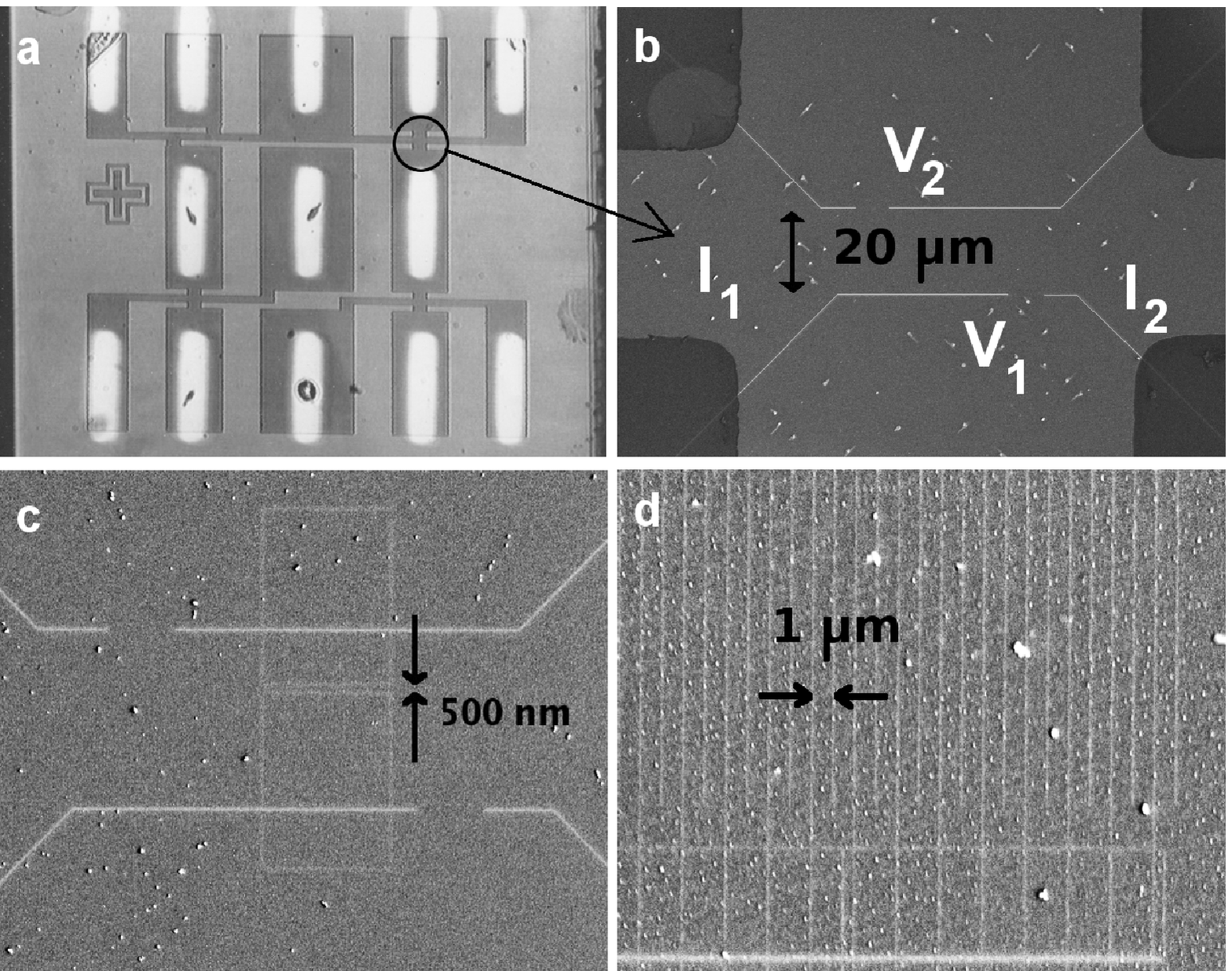}

\caption{
\label{fig:all_sem}
Overview of the patterning protocol. \mbox{({\em{a}}) Photograph} of the preliminary structure created from a thin film by photolithography and chemical etching. \mbox{({\em{b}}) SEM} micrograph of a 20 $\mu$m stripe engineered with a high-current Focused Ion Beam. \mbox{({\em{c}}) SEM} micrograph of a 12 nm thick, 15 $\mu$m long, 500 nm wide stripe patterned with a lower current. \mbox{({\em{d}}) SEM} micrograph of a 1 $\mu$m wide meandering circuit written with a lower current.
} 
\end{figure}

We designed a 2-step protocol, involving a preliminary chemical etching followed by a focused ion beam (FIB) managed nanostructuration. An overview of the scheme used to create YBCO circuits embedding paths to characterize their transport properties with 4-point measurements is given in figure \ref{fig:all_sem}.

\subsection{Growth of YBCO films}

\begin{figure}[p]
\includegraphics[width=\fgwidth]{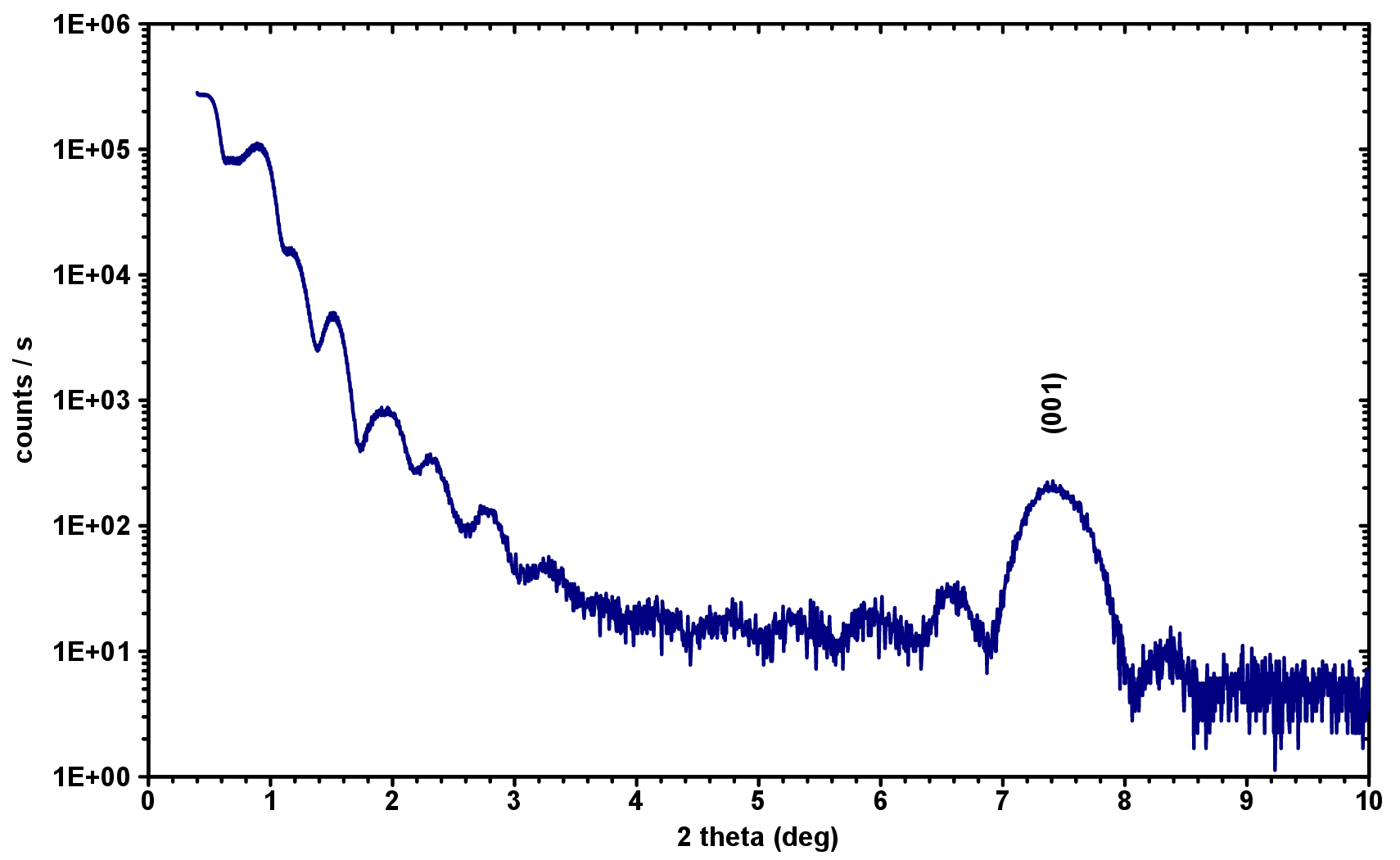}

\caption{
\label{fig:ybco_thickness}
Typical X-ray $\theta$-2$\theta$-spectrum of a 12 nm thick YBCO film passivated with a 8 nm thick amorphous PBCO layer.
}
\end{figure}

$c$-axis YBCO films were deposited by RF magnetron sputtering on (100) SrTiO$_\text{3}$ substrates. Samples are heated to 700°C and exposed to the sputtering of a YBCO target in a Argon plasma. Along with the Ar flow a complementary O$_\text{2}$ flow (1:5 ratio) accounts for the growth of a tetragonal, non-superconducting phase, under a total pressure P=8.10$^\text{-2}$ mbar. An in-situ 2-hour-long annealing at 580°C in a \mbox{600 mbar} O$_\text{2}$ atmosphere causes this YBa$_\text{2}$Cu$_\text{3}$O$_{6}$ phase to undergo a tetragonal-orthorhombic transition to optimally-doped superconducting YBa$_\text{2}$Cu$_\text{3}$O$_{\text{7-}\delta}$. The critical temperature of bulk YBCO is 92 K; this critical temperature decreases with the film's thickness, down to T$_\text{c0}$(d=12 nm) $\approx$ 80 K. The high crystallinity of the films was demonstrated with X-ray diffraction measurements such as depicted in figure \ref{fig:ybco_thickness}. Up to 4 degrees the grazing incidence scan leads to Kiessig fringes contributions from both films \cite{Stoev-SAB-1999}. Around the (001) YBCO Bragg peak are located secondary fringes clearly showing a finite size effect (Laue oscillations), demonstrating the high quality of the crystallographic layers. The fringes of X-ray spectra allow to determine a film's thickness with unit cell accuracy. In the following, all the films processed are at d=12 nm, with T$_\text{c0} \approx$ 80 K.

The sputtering process also deposits CuO$_\text{2}$ surface particles located on the top of the YBCO phase, with a typical diameter estimated at 250 nm; those might indeed be the origin of shortcuts after patterning. No reproducible way of getting particle-free samples or eliminating them has been found, however most of the time these particles are not a real obstacle to the patterning process, since the core structure of the devices can most of the times be chosen to be located in a clear area (figure \ref{fig:all_sem}c). Even when it's not the case, such as in figure \ref{fig:all_sem}d, table \ref{tab:repro} shows that ultimately these particles are not an issue.

In order to improve electrical contacts, we adapted \vtwo{a} gold evaporation system involving a mechanical mask to deposit gold slots in-situ immediatly after the YBCO deposition. In addition it was observed that the in-situ deposition of a 8 nm thick amorphous PrBaCuO passivation cap layer over the whole sample attenuates the loss of critical temperature occurring during the chemical etching process \cite{Jaeger-ITAS-1993}. Therefore, we end up with a YBCO/Au/PBCO topology.

\subsection{Photolithography and chemical etching}

The patterning of a 50 $\mu$m wide stripe is done by optical lithography and chemical etching with orthophosphoric acid H$_\text{3}$PO$_\text{4}$ (1\%). Four independent structures are patterned in a single run (figure \ref{fig:all_sem}a). One of them, top-left on the figure, is specifically designed to electrically characterize the sample at this point of the protocol.

\subsection{FIB writing}

As the second step of the protocol, a dual beam (FEI Nova 600 Nanolab scanning electron microscope / 30 kV focused Ga$^\text{3+}$ ion beam) was used to carry out a finer patterning of the YBCO films into meandering stripes suitable for optical characterization of SSPDs. In the context of YBCO thin layers, some precautions to make a relevant use of the FIB are necessary since the standard manner of working is too destructive for the samples. Accordingly, we devised a specific modus operandi to satisfy the needed requirements.

First of all, due to the extreme thickness of the involved films, the alignment routines cannot be handled with the standard procedure, as the whole window, containing areas destined to remain superconducting is exposed to the Ga$^\text{3+}$ beam during the operation, and irreversibly damaged. To circumvent this problem we have to synchronize the SEM and the FIB on a non-critical area, then align the pattern to etch with the sample using the SEM. 

\vtwo{
Secondly, to find which parameters (intensity, number of passes, numerical aperture...) are optimal to produce an effective pattern, we tried to separate two YBCO areas by creating FIB-made barriers and injecting current through them; results for a 50 pA beam are shown in table \ref{tab:npass}. We determined that at 50 pA, with a beam diameter $\delta\approx$ 20 nm, 50 passes are optimal to electrically isolate both areas, while at 1 nA, with a beam diameter $\delta\approx$ 100 nm, 10 passes are optimal. The fact that for N>20 the resistance exceeds 3 M$\Omega$ rules out an exclusive YBCO$_\text{7}$ to YBCO$_\text{6}$ transformation due to oxygen loss, for if it were the case, the barrier transition would have resulted in a 5 k$\Omega$ resistance (assuming very conservatively $\rho$(underdoped)=10$^\text{-1}$ $\Omega$.cm at room temperature \cite{Semba-PHYSB-2000}) for a 1 $\mu$m wide affected width. Even if the whole stripe had been transformed to YBCO$_\text{6}$ its resistance wouldn't have exceeded 150 k$\Omega$. We attribute this result to the fact that some Ga$^\text{3+}$ ions are implanted or give rise to columnar defects into the YBCO phase during the exposure, turning it locally into an insulator.
}

\begin{table}
\vtwo{
\caption{\label{tab:npass}Room-temperature resistance of a FIB-made barrier across a 20 $\mu$m wide stripe for various numbers of passes (N) with a 50 pA beam.}
\begin{indented}
\item[]\begin{tabular}{@{}cccccccc}
\br
N & 0 & 1 & 2 & 5 & 20 & 50 & 100 \\
\mr
R ($\Omega$) & 2k  & 2k & 3k & 700k & 3M &>10M &>10M \\
\br
\end{tabular}
\end{indented}
}
\end{table}

\vtwo{Lateral contamination is a crucial parameter to consider since it directly addresses both the issues of smallest reachable dimensions and current homogeneity. To illustrate this point, we obtained different kinds of superconducting transitions on structures of different widths by varying the number of FIB passes N. Figure \ref{fig:fib_effect} shows a comparison between resistive profiles for 20 $\mu$m wide stripes patterned with respectively N=500 and N=10 passes. A significant loss of T$_\text{c}$ occurs for N=500. This T$_\text{c}$ loss may be due to a reduction of the O$_\text{2}$ stoichiometry, but in this case it can't be attributed to local heating occurring during the writing process. Indeed, the elevation of the film temperature has been estimated within a simple thermal model and found to be lower than 5 K, under the beam writing conditions (voltage: 30 kV, current: 1 nA, diameter: 100 nm). Besides, it is unlikely that Ga ions can travel 10 $\mu$m accross the sample resulting in implantations responsible for the T$_\text{c}$ losses observed in figure \ref{fig:fib_effect}. Consequently N=500 was ruled out for our process. For N=10 no such loss happens, although the current is confined in the stripe. 

\begin{figure}[p]
\includegraphics[width=\fgwidth]{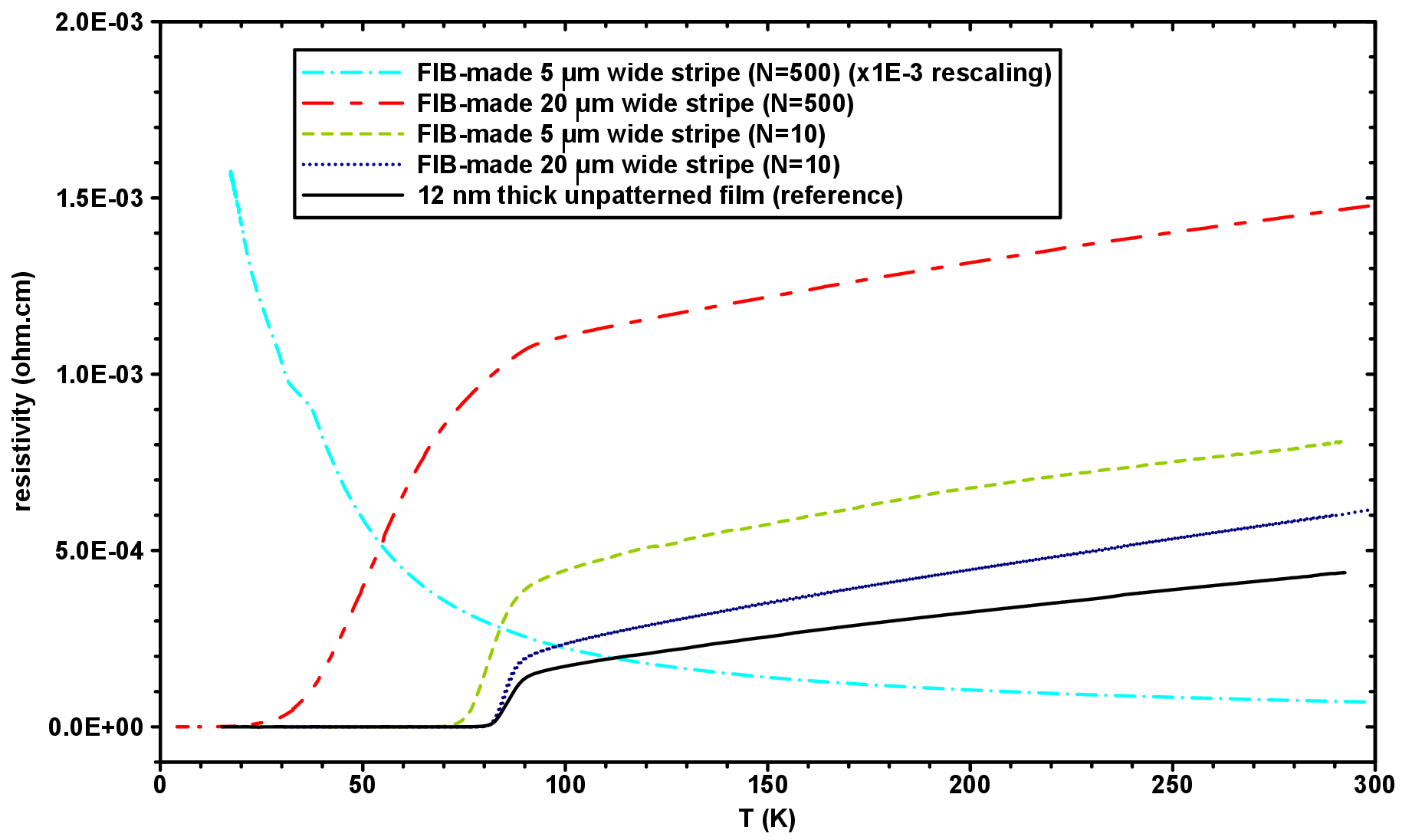}

\caption{
\label{fig:fib_effect}
Resistive curves of 12 nm thick micron-scale circuits patterned using a Focused Ion Beam with a beam current i=1 nA. N is the number of passes used to produce the structure.}
\end{figure}

Finally, to pattern a meandering wire, we first produce a 20 $\mu$m wide stripe (figure \ref{fig:all_sem}b), then we force the current to follow a serpentine path by adding some crossing barriers (figure \ref{fig:all_sem}d). The supporting stripe has to be manufactured at 1nA because of window size considerations. Thus we choose to use two different FIB currents to pattern a meander : 1 nA, 10  passes for the supporting stripe; 50 pA, 50 passes for the meander itself. Supplementary barriers parallel to the stripe are added to avoid side effects near to the 1nA-exposed areas. Using this protocol superconducting straight stripes down to 500 nm wide (figure \ref{fig:all_sem}c), and meanders down to 1 $\mu$m wide (figure \ref{fig:all_sem}d) were produced. It's worth mentioning that at no point we were confronted to some limitation so in principle it should be possible to achieve even better resolutions with this method.
}

\section{Results and discussion}

\begin{figure}[p]
\includegraphics[width=\fgwidth]{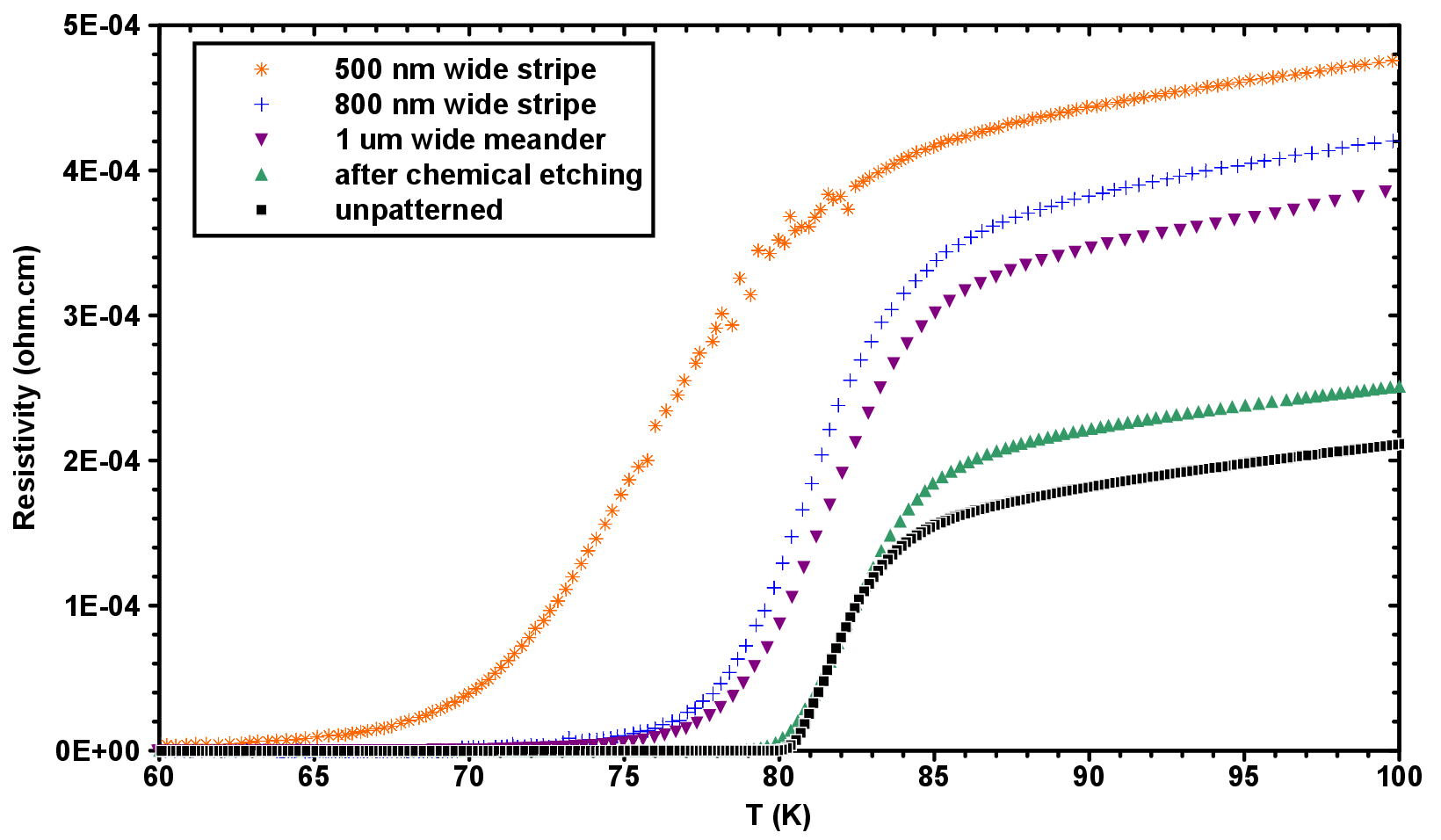}

\caption{
\label{fig:rvst}
Resistivity vs temperature for different samples, with additional curves at intermediary steps of the process.
}
\end{figure}

The $\rho$ vs T curves of the samples can be followed during the different steps of the protocol: after the film deposition the resistivity is measured with the Van der Pauw method \cite{VanderPauw-PTR-1958}; after chemical etching a four-point measurement is carried out along a 50 $\mu$m wide stripe using the top-left structure of figure \ref{fig:all_sem}a. Eventually the final circuit is characterized as shown in figure \ref{fig:all_sem}b. Results are plotted in figure \ref{fig:rvst}, and demonstrate that the obtained circuits are superconducting, although a small loss of T$_\text{c}$ as well as a broadening of the resistive transition is observed. This could be explained by the fact that the samples are intrinsically inhomogeneous, and by restricting the superconducting geometry to narrow circuits new areas with lower T$_\text{c}$ enter the current path; but it could also be material damages occuring during the processing. On another hand, $\rho_\text{100K}$ and $\rho_\text{300K}$ are slightly higher at the end of the process.

As mentioned previously the writing routine with the focused ion beam doesn't etch the film; instead the beam turns locally the irradiated area into an insulating phase by implanting columnar defaults inside it. The penetration depth of Ga$^\text{3+}$ ions is about 70 nm, far superior to the YBCO and amorphous PBCO combined thickness. From this point of view this method differs from EBL/IBE processes where parts of the film are physically removed to create the superconducting pattern, but it also implies caution if one desires to implement it on with thicker films. This approach prevents the YBCO phase from being in contact with air, which could be an escape path for oxygen, both with the PBCO top passivation layer and the lateral insulating phase. Table \ref{tab:repro} presents resistance measurements obtained for different structures, the last column showing the consistency of the results and ensuring that the method presented in the paper leads to reproducible structures.

\begin{table}
\caption{\label{tab:repro}Summary of T$_\text{c}$ and resistance for different kinds of samples. R$_\text{norm}$ is the equivalent resistance computed with the structure normalized to a 1 $\mu$m wide, 15 $\mu$m long line.}
\begin{indented}
\item[]\begin{tabular}{@{}cccccc}
\br
shape & length & width & T$_\text{c0}$ & R(100K) & R$_\text{norm}$ ($k\Omega$)\\
\mr
meander & 180 $\mu$m & 2 $\mu$m & 72 K & 32 k$\Omega$ & 5.3 \\
meander & 430 $\mu$m & 1 $\mu$m & 70 K & 190 k$\Omega$ & 6.6 \\
meander & 430 $\mu$m & 1 $\mu$m & 75 K & 140 k$\Omega$ & 4.9 \\
stripe & 15 $\mu$m & 1 $\mu$m & 70 K & 5 k$\Omega$ & 5 \\
stripe & 15 $\mu$m & .8 $\mu$m & 75 K & 6.6 k$\Omega$ & 5.3 \\
stripe & 15 $\mu$m & .5 $\mu$m & 65 K & 12 k$\Omega$ & 6 \\
\br
\end{tabular}
\end{indented}
\end{table}

\begin{figure}[p]
\includegraphics[width=\fgwidth]{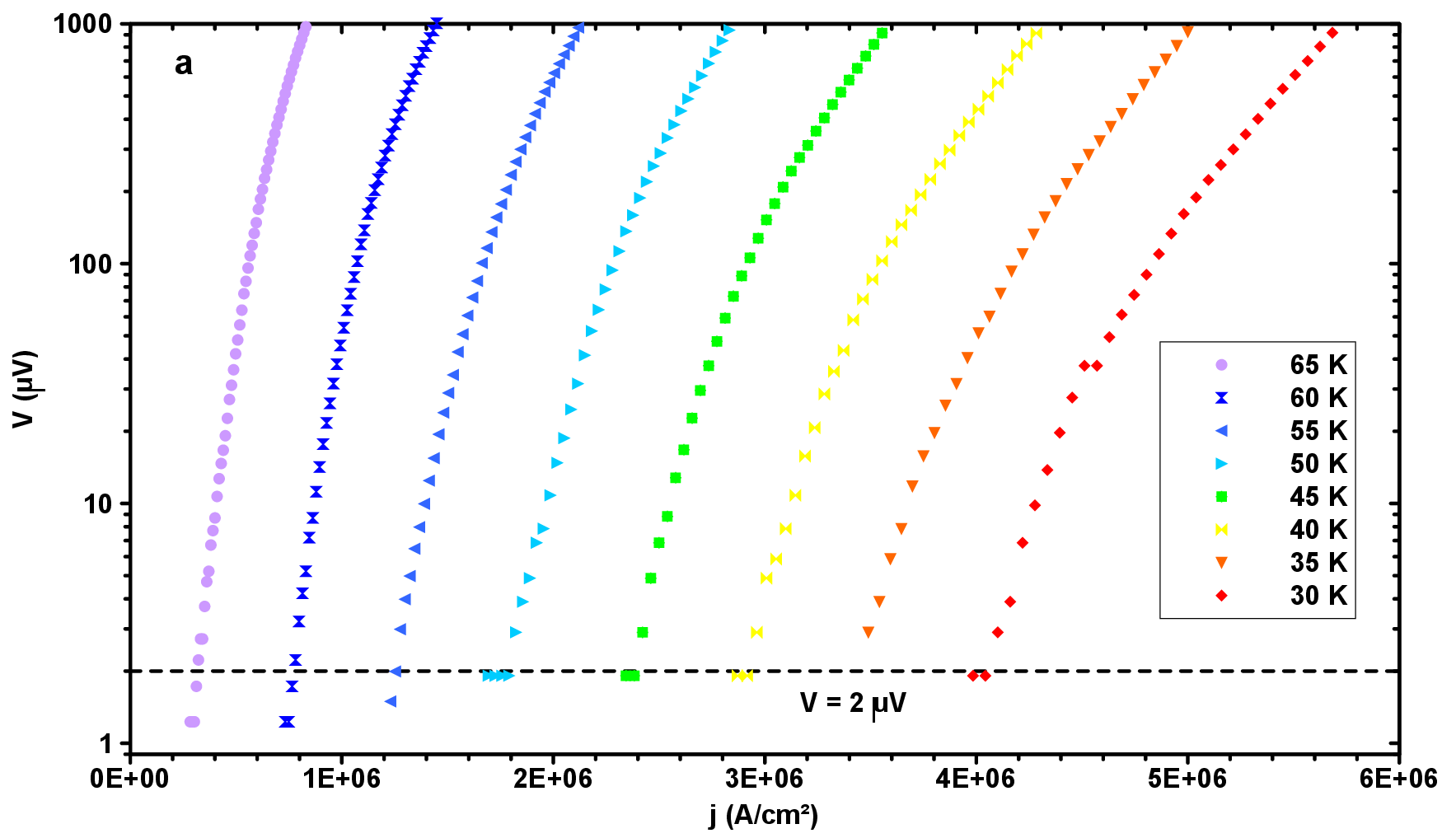}
\includegraphics[width=\fgwidth]{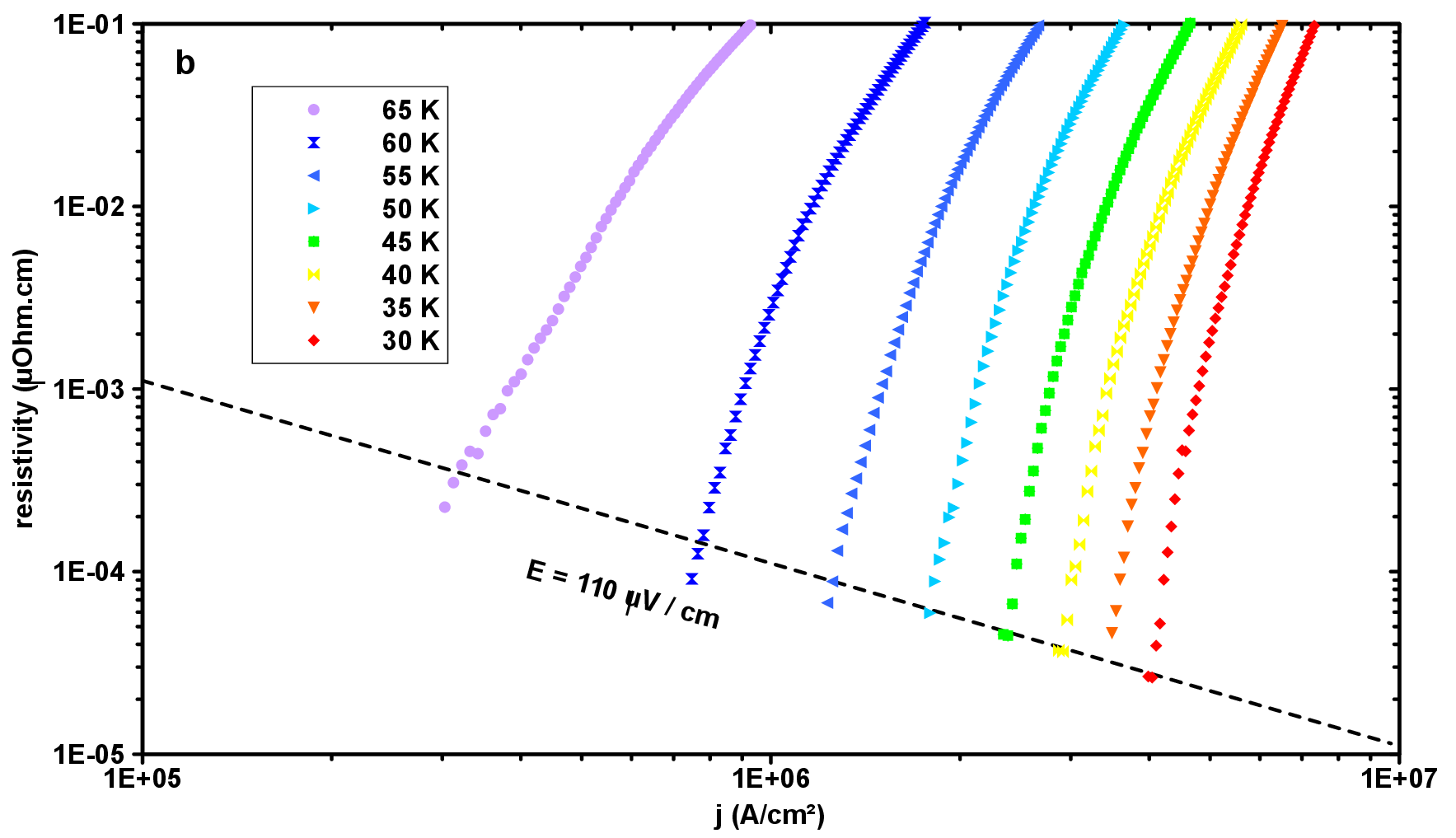}

\caption{
\label{fig:rhovsj}
({\em{a}}) Voltage vs current density at various temperatures for a 2 $\mu$m wide meandering circuit. \mbox{({\em{b}}) Resistivity} vs current density for the same sample.
}
\end{figure}

Figure \ref{fig:rhovsj} present the current-voltage and the $\rho$-$j$ characteristics of one sample. Both present a negative curvature in the whole range of temperature covered, yielding the existence, in every case, of a true critical current density $j_\text{c}$. This ensures that the superconducting phase isn't in a flux-creep state in spite of the thinness of the sample, which is consistent with the observation that above \mbox{$d$ = 10 nm} flux line lattices behave like a 3D system \cite{Triscone-REPRO-1997}. 

The critical current is generally defined using the standard criterion of an electric field of 1$\mu$V/cm. In our case the measurement noise due to the high impedance of the line sets a resolution limit of 2 $\mu$V across the whole meander, corresponding to an equivalent electric field \mbox{E = 110 $\mu$V/cm} as shown on figures \ref{fig:rhovsj}a and \ref{fig:rhovsj}b. We clearly see on figure \ref{fig:rhovsj}b, especially at T = 65 K, that at 110 $\mu$V/cm (or 2 $\mu$V) the critical current densities are overestimated and that such a determination isn't relevant.

\begin{figure}[p]
\makebox[\mbl][l]{\includegraphics[width=\fgwidth]{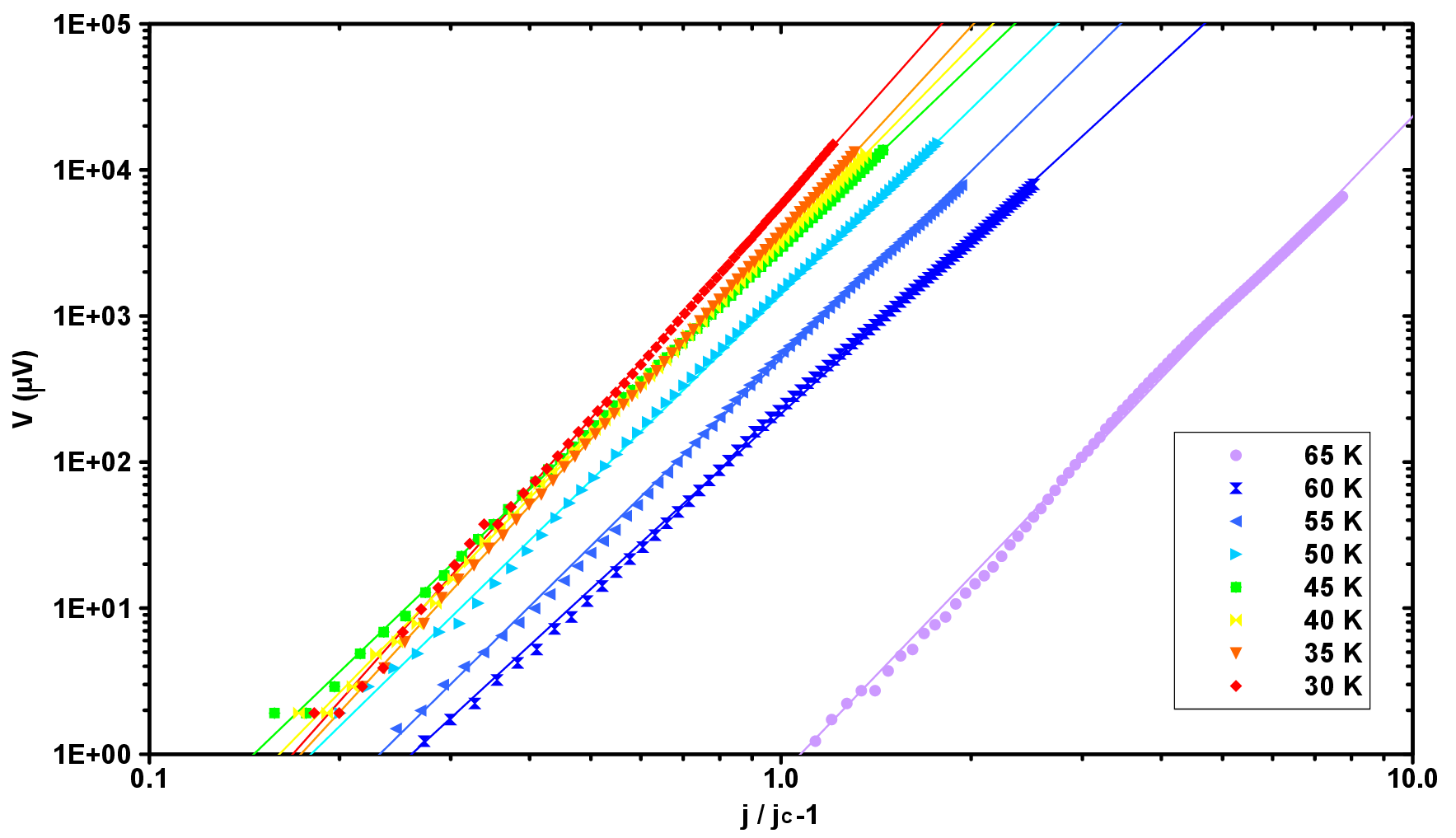}}
\raisebox{\rbl}[0pt][0pt]{\includegraphics[width=\iswidth]{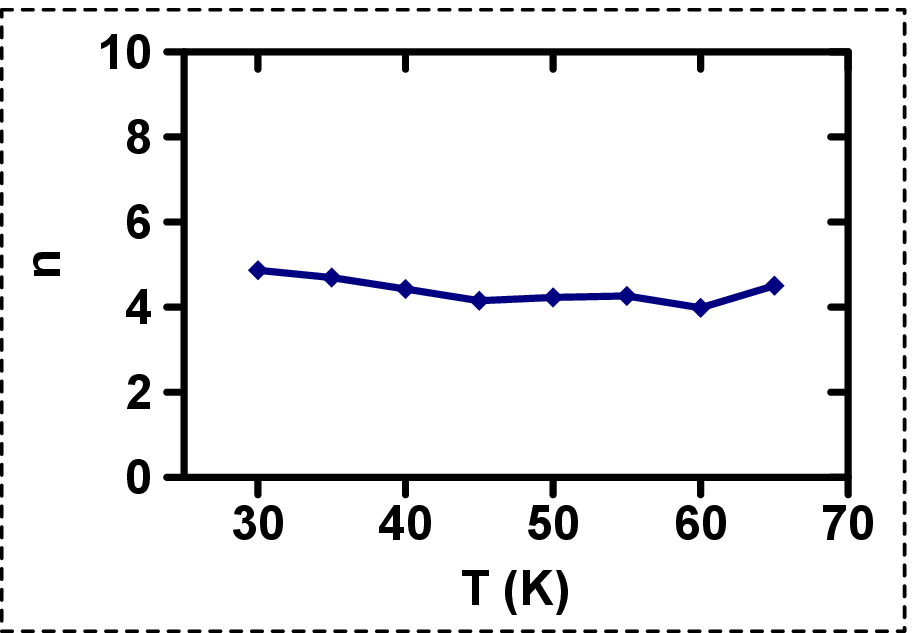}}

\caption{
\label{fig:jc}
Same data as in figure \ref{fig:rhovsj}a represented as voltage vs $j/j_\text{c}-1$. Power law fits $V=B(j-j_\text{c})^n$ (continuous lines) correctly describe the curves and allow to determine $j_\text{c}$. In inset is reported $n$ as temperature drops from 65 K to 30 K.
}
\end{figure}

However, it has been reported \cite{Antognazza-PHYSC-2002} that a precise measurement of the flux flow resistivity at high current density allows to extract $j_\text{c}$ from the best power law fit \mbox{$V\propto(j-j_\text{c})^n$}, $j_\text{c}$ and $n$ being the fitting parameters. Figure \ref{fig:jc} shows the very good fits obtained with power laws, giving good confidence for the determination of $j_\text{c}$'s by this procedure. Moreover, $n$ is found to be around 5 with little variation over the whole temperature range. A significant advantage of this fitting method is that the knowledge of the current-voltage characteristic at high current density is sufficient to infer $j_\text{c}$.

\begin{figure}[p]
\includegraphics[width=\fgwidth]{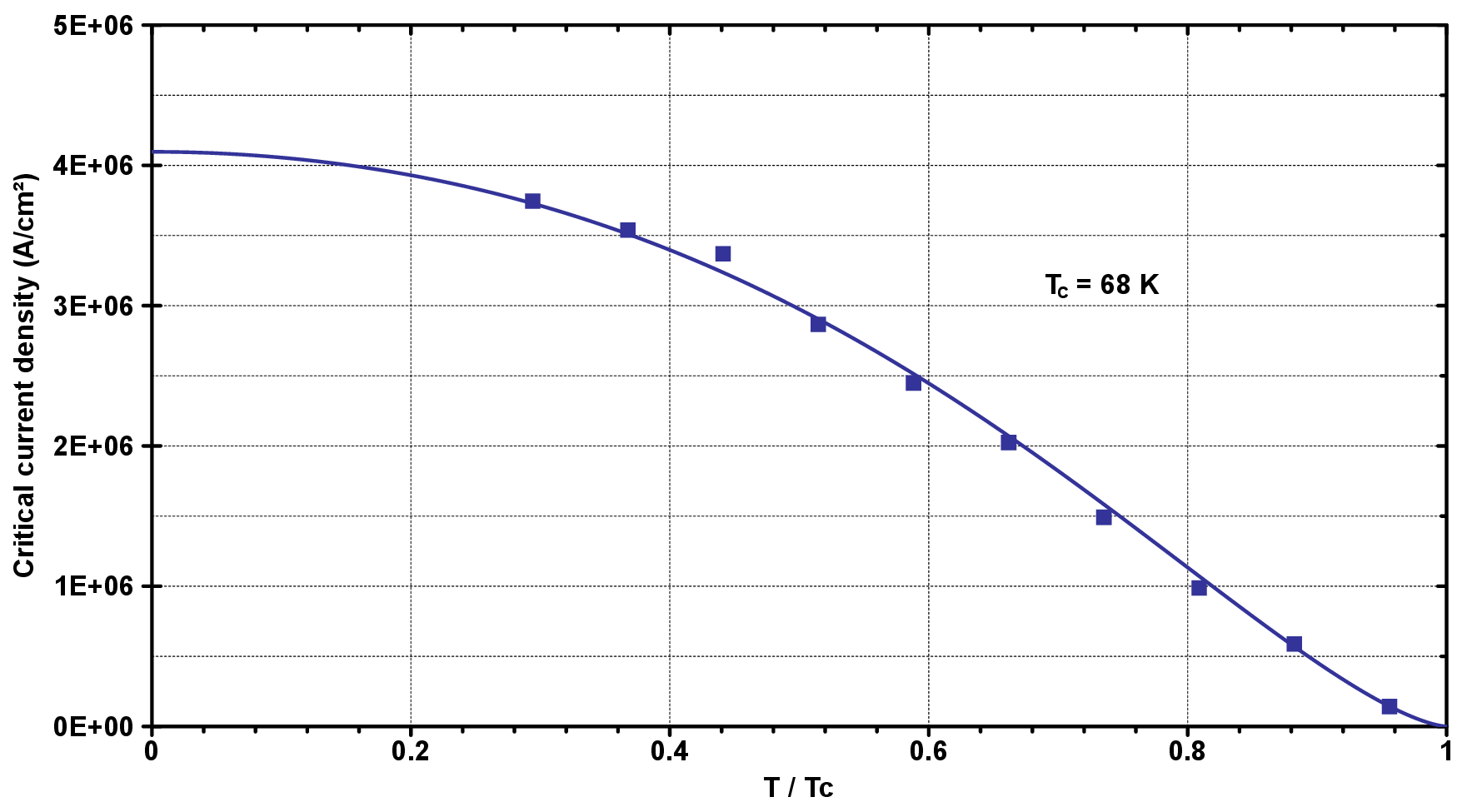}

\caption{
\label{fig:gl}
Critical current density vs reduced temperature $t=T/T_\text{c}$ for the same meandering circuit. $j_\text{c}$ is obtained through the power law fits presented in figure \ref{fig:jc}. The solid line is the best fit with the theoretical Ginzburg-Landau model. 
}
\end{figure}

The temperature-dependence of $j_\text{c}$ is described within the Ginzburg-Landau theory of superconductivity as \mbox{$j_\text{c}$ = $j_\text{c0} (1-(\frac{T}{T_\text{c}})^2) (1-(\frac{T}{T_\text{c}})^4)^{1/2}$} \cite{Poole-Super-1995}. Figure \ref{fig:gl} shows that this model fits extremely well our experimental data and allows to extrapolate \mbox{$j_c(0K)\approx$ 4.1 MA/cm$^\text{2}$}, which is two orders of magnitude smaller than the depairing limit \mbox{$j_\text{d}=0.54\frac{B_\text{c}}{\mu_0 \lambda}\approx 3~10^{8}$ A/cm$^\text{2}$}, demonstrating the good quality of the samples after the processing. It's worth noting that the hypothesis of inhomogeneities in the sample previously mentioned is reinforced by the fact that the best Ginzburg-Landau fit is obtained for \mbox{T$_\text{c}$ = 68 K}, below the critical temperature of 72 K found with resistivity measurements (see table \ref{tab:repro}).

\vtwo{A last point concerns long-term behavior of the nanowires. Although no systematic study was carried out, some samples were characterized several times at low temperature, with more than one month gone by between the measurements. No deterioration of any kind was observed, which is an indication that the manufactured structures are unaffected by time and thermal cycling.}

\section{Conclusion}

To summarize, a modus operandi to create ultrathin superconducting YBCO circuits by implanting Ga$^\text{3+}$ ions with a Focused Ion Beam was devised. We confined current in straight stripes down to 500 nm wide and produced 1 $\mu$m wide meandering wires using a 50 pA beam current. The consistency of the resistivity vs temperature profiles measured on the samples at the different steps of the processing ensures the reproducibility of this patterning method for superconducting films. For one sample, the critical current density extrapolated to 0 K has been found to be only two orders of magnitude times smaller than the depairing limit, demonstrating its quality. The most natural application of this protocol would be the manufacturing of high-T$_\text{c}$ superconducting devices. Photoresponse experiments to characterize the devices as single-photon detectors are left for future work.

\ack

The authors would like to thank Michaël Pavius, Kevin Lister, Samuel Clabecq, and Philippe Flückiger for providing access and training to EPFL's focused ion beam, as well as Jean-Claude Villégier for helpful discussions. This work is supported by the European project Sinphonia (contract No. NMP4-CT-2005-16433), and the Swiss poles NCCR MaNEP and NCCR Quantum Photonics.

\bibliographystyle{unsrt}
\bibliography{article}

\end{document}